Hitting Streaks Don't Obey Your Rules
Evidence That hitting Streaks Aren't Just By-products of Random Variations

Trent McCotter, UNC Chapel Hill, Jan 2009, contact: trentm@email.unc.edu .

Professional athletes naturally experience hot and cold streaks.  However, there's been a debate going on for some time now as to whether professional athletes experience streaks more frequently than we would expect given the players' season statistics.  This is also known as having "the hot hand."

For example, if a player is a 75 percent free-throw shooter this season and he's made his last 10 free throws in a row, does he still have just a 75 percent chance of making the 11thfree throw? The answer from most statisticians would be a resounding *Yes*, but many casual observers believe that the player is more likely to make the 11th attempt because he's been "hot" lately and that his success should continue at a higher rate than expected. Two common explanations for why a player may be "hot" are that his confidence is boosted by his recent success or that his muscle memory is better than usual, producing more consistency in his shot or swing.

**As it relates to baseball**

The question is this: Does a player's performance in one game (a '"trial," if you will) have any predictive power for how he will do in the next game (the next trial)? If a baseball player usually has a 75 percent chance of getting at least one base hit in any given game and he's gotten a hit in 10 straight games, does he still have a 75 percent chance of getting a hit in the 11th game? This is essentially asking, "Are batters' games independent from one another?"

As with the free-throw example, most statisticians will say that the batter in fact does still have a 75 percent of getting a hit in the next game, regardless of what he did in the last 10. In fact, this assumption has been the basis for several *Baseball Research Journal* articles in which the authors have attempted to calculate the probabilities of long hitting streaks, usually Joe DiMaggio's major-league record 56-game streak in 1941. It was this assumption about independence that I wanted to test, especially in those rare cases where a player has a long hitting streak (20 consecutive games or more). These are the cases where the players are usually aware that they've got a long streak going.

If it's true that batters who are in the midst of a long hitting streak will tend to be more likely to continue the streak than they normally would (they're on a "hot streak"), then we would expect more 20-game hitting streaks to have *actually* happened than we would *theoretically* expect to have happened. That is, if players realize they've got a long streak going, they may change their behavior (maybe by taking fewer walks or going for more singles as opposed to doubles) to try to extend their streaks; or maybe they really are in an abnormal 'hot streak.' But how do we determine what the theoretical number of twenty-game hitting streaks should be?

In the standard method, we start by figuring out the odds of a batter going hitless in a particular game, and then we subtract that value from 1; that will yield a player's theoretical probability of getting at least one hit in any given game:

$$1-((1-(AVG))^{(AB/G)})$$

For a fabricated player named John Dice who hit .300 in 100 games with 400 at-bats, this number would be:

$$1-((1-(.300))^{(400/100)}) = .7599 = 76 \text{ percent chance of at least one hit in any given game}$$

With the help of RetroSheet's Tom Ruane, I did a study over the 1957–2006 seasons to see how well that formula can predict the number of games in which a player will get a base hit. For example, in the scenario above, we would expect John Dice to get a hit in about 76 of his games; it turns out the formula above is indeed very accurate at predicting a player's *number* of games with at least one hit.

Thus, if games really are independent from one another and don't have predictive power when it comes to long hitting streaks, this means that John Dice's 100-game season can be seen as a series of 100 tosses of a weighted coin that will come up heads 76 percent of the time; long streaks of heads will represent long streaks of getting a hit in each game. This method for calculating the odds of hitting streaks was used by Michael Freiman in his article "56-Game Hitting Streaks Revisited" in *BRJ* 31 (2002), and it was also used by the authors of a 2008 op-ed piece in the New York Times:

> Think of baseball players' performances at bat as being like coin tosses. Hitting streaks are like runs of many heads in a row. Suppose a hypothetical player named Joe Coin had a 50–50 chance of getting at least one hit per game, and suppose that he played 154 games during the 1941 season. We could learn something about Coin's chances of having a 56-game hitting streak in 1941 by flipping a real coin 154 times, recording the series of heads and tails, and observing what his longest streak of heads happened to be.
>
> Our simulations did something very much like this, except instead of a coin, we used random numbers generated by a computer. Also, instead of assuming that a player has a 50 percent chance of hitting successfully in each game, we used baseball statistics to calculate each player's odds, as determined by his actual batting performance in a given year.
>
> For example, in 1941 Joe DiMaggio had an 81 percent chance of getting at least one hit in each game . . . we simulated a mock version of his 1941 season, using the computer equivalent of a trick coin that comes up heads 81 percent of the time.
>
> —Samuel Arbesman and Steven Strogatz, *New York Times*, 30 March 2008

But I wondered whether this method has a fundamental problem as it relates to looking at long hitting streaks, because it uses a player's *overall* season stats to make inferences about what his season must have looked like on a *game-by-game* basis.

Think of the example of flipping a coin. That's about as random as you can get, and we wouldn't really consider the outcome of your last flip to affect the outcome of your next flip. That means that we can rearrange those heads and tails in any random fashion and the only variation in streaks of heads would be due entirely to random chance. *If this were true in the baseball example*, it means that we could randomly rearrange a player's season game log (listing his batting line for each game) and the only variation in the number of long streaks that we would find would be due entirely to random chance.

**The Number-Crunching**

To see who's right about this, we need to solve the problem of how to calculate the theoretical number of hitting streaks we would expect to find. It turns out that the answer actually isn't too complicated. I took the batting lines of all players for 1957 through 2006 and subtracted out the 0-for-0 batting lines, which neither extend nor break a hitting streak. I ended up with about 2 million batting lines.

Then, with the impressive assistance of Dr. Peter Mucha of the Mathematics Department at the University of North Carolina, I took each player's game log for each season of their career and sorted the game-by-game stats in a completely random fashion. So this means that, for instance, I'm still looking at John Dice's .300 average, 100 games, and 400 at-bats—but the order of the games isn't chronological anymore. It's completely random. It's exactly analogous to taking the coin tosses and sorting them randomly over and over to see what long streaks of heads will occur. See the example at the end of this article for a visual version of this.

Dr. Mucha and I ran each random sorting *ten thousand* separate times, so we ended up sorting every player-season from 1957 through 2006 ten thousand separate times to see what streaks occurred. For each of the 10,000 permutations, we counted how many hitting streaks of each length occurred. The difference between this method and the method that has been employed in the past is that, by using the actual game-by-game stats (sorted randomly for each player), we don't have to make theoretical guesses about how a player's hits are distributed throughout the season. Remember, if players' games were independent from one another, this method of randomly sorting each player's games should—in the long run—yield the same number of hitting streaks of each length that happened in real life.

Here are the results.

| 1957–2006 | **Actually Happened** | **In the 10,000 Random Sortings** | |
|---|---|---|---|
| Length | Number of Hitting Streaks | Average | Standard Deviation |
| 5 | 22,632 | 22,584.63 | 141 |
| 6 | 14,470 | 14,086.60 | 112 |
| 7 | 9,151 | 8,947.64 | 89 |
| 8 | 6,081 | 5,766.29 | 72 |
| 9 | 4,059 | 3,759.81 | 59 |
| 10 | 2,645 | 2,477.50 | 48 |
| 11 | 1,792 | 1,647.42 | 39 |
| 12 | 1,226 | 1,104.86 | 32 |
| 13 | 801 | 747.12 | 27 |
| 14 | 552 | 506.85 | 22 |
| 15 | 415 | 347.13 | 19 |
| 16 | 270 | 238.69 | 16 |
| 17 | 194 | 164.97 | 13 |
| 18 | 129 | 114.22 | 11 |
| 19 | 112 | 79.80 | 8.9 |
| 20 | 75 | 55.90 | 7.5 |
| 21 | 52 | 39.36 | 6.2 |
| 22 | 38 | 27.80 | 5.3 |
| 23 | 25 | 19.70 | 4.4 |
| 24 | 22 | 13.93 | 3.7 |
| 25 | 17 | 10.00 | 3.2 |
| 26 | 8 | 7.20 | 2.7 |
| 27 | 7 | 5.13 | 2.3 |
| 28 | 7 | 3.71 | 1.9 |
| 29 | 4 | 2.63 | 1.6 |
| 30 | 9 | 1.90 | 1.4 |
| 31 | 4 | 1.39 | 1.2 |
| 32 | 0 | 1.01 | 0.99 |
| 33 | 0 | 0.74 | 0.86 |
| 34 | 1 | 0.55 | 0.74 |
| 35+ | 5 | 1.48 | 1.21 |

| 1957–2006 | Actually Happened | In The 10,000 Random Sortings | |
|---|---|---|---|
| Length | Number of Hitting Streaks | Average | Standard Deviation |
| 5+ | 64,803 | 62,765.96 | 150.69 |
| 10+ | 8,410 | 7,620.99 | 69.45 |
| 15+ | 1,394 | 1,137.24 | 30.71 |
| 20+ | 274 | 192.43 | 13.32 |
| 25+ | 62 | 35.74 | 5.84 |
| 30+ | 19 | 7.07 | 2.60 |
| 35+ | 5 | 1.48 | 1.21 |

It's clear that, for every length of hitting streak of 5-plus games, there have been more streaks in reality than we would expect given players' game-by-game stats. To give those numbers some meaning: There were 19 single-season hitting streaks of 30-plus games from 1957 through 2006. The ten thousand separate, random sortings of the game-by-game stats produced an average of 7.07 such streaks for 1957–2006. That means that almost *three times* as many 30-plus-game hitting streaks have occurred in real life as we would have expected.

Since there were 10,000 trials for our permutation, the numbers here are all highly significant. For instance, the average number of 5-plus-game streaks in the permutations was about 62,766, with a standard deviation of about 151, and there were 64,803 such streaks in real life from 1957 through 2006. This means that the real-life total was 13.5 deviations away from the expected mean, which implies that the odds of getting these numbers simply by chance are about *one in 150 duodecillion* (150 followed by 39 zeros). The number of hitting streaks that have really happened is significantly much higher than we would expect if long hitting streaks could in fact be predicted using the coin-flip model. Additionally, the results of the 10,000 trials converged, which means that the first 5,000 trials had almost the exact same averages and standard deviations as did the second 5,000 trials.

But what does this all mean? What it seems to indicate is that many of the attempts to calculate the probabilities of long hitting streaks are actually underestimating the true odds that such streaks will occur. Additionally, if hits are not IID (independent and identically distributed) events, then it may be extremely difficult to devise a way to calculate probabilities that do produce more accurate numbers.

**So why don't the permutations match the real-life numbers?**

It's easier to begin by debunking several common-sense explanations as to why the permutations didn't produce a similar number of hitting streaks as happened in real life.

The first one I thought of was the quality of the opposing pitching. If a batter faces a bad pitching staff, he'd naturally be more likely to start or continue a hitting streak, relative to his overall season numbers. But the problem with this explanation is that it's too short-sided; you can't face bad pitching for too long without it noticeably increasing your numbers, plus you can't play twenty games in a row against bad pitching staffs, which is what would be required to put together a long streak. This same reasoning is why playing at a hitter-friendly stadium doesn't seem to work either, since these effects don't continue for the necessary several weeks in a row. In other words, the explanation must be something that lasts longer than a four-game road trip to Coors Field or getting to face Jose Lima twice in one month.

The second possible explanation—one that I really thought could explain everything—was the weather. That is, it's commonly believed that hitting increases during the warm summer months, which would naturally make long hitting streaks more likely, while the cooler weather at the beginning and the end of the season makes streaks less likely. This would explain why long streaks seem to happen so much more often than we'd expect; the warmest period of the summer can last for months, seemingly making it fertile ground to start a hitting streak. The reason this is important is that hitting streaks are exponential. That is, a player who hits .300 for two months will be *less* likely to have a hitting streak than a player who hits .200 one month and .400 the next; the two players would have the exact same numbers, but hitting streaks tend to highly favor batters who are hitting very well, even if it's just for a short period, and even if it's counterbalanced by a period of poor hitting.

The problem with the weather explanation is that the stats don't bear it out. Of the 274 streaks of 20-plus games from 1957 through 2006, there were just as many that began in May as began in June, July, or August. If it were true that the hottest months spawned hitting streaks, we would see a spike in streaks that began in those months. We don't see that spike at all.

So that eliminates the explanations that would seem to be the most likely. Remember, if all of the assumptions about independence were right, we wouldn't even have these differences between the expected and actual number of streaks in the first place; so it's yet another big surprise that our top explanations for these discrepancies also don't seem to pan out. This leaves me with two other possible explanations, each of which may involve psychology more than mathematics.

**First explanation**

Maybe the players who have long streaks going will change their approach at the plate and go for fewer walks and more singles to keep their hitting streaks going. This same idea is covered in *The Bill James Goldmine*, where James discusses how pitchers will make an extra effort to reach their 20th victory of a season, which results in there being more 20-game winners in the majors than 19-game winners. There is evidence of this effect, too, as seen by the following graph, which visualizes the chart from earlier.

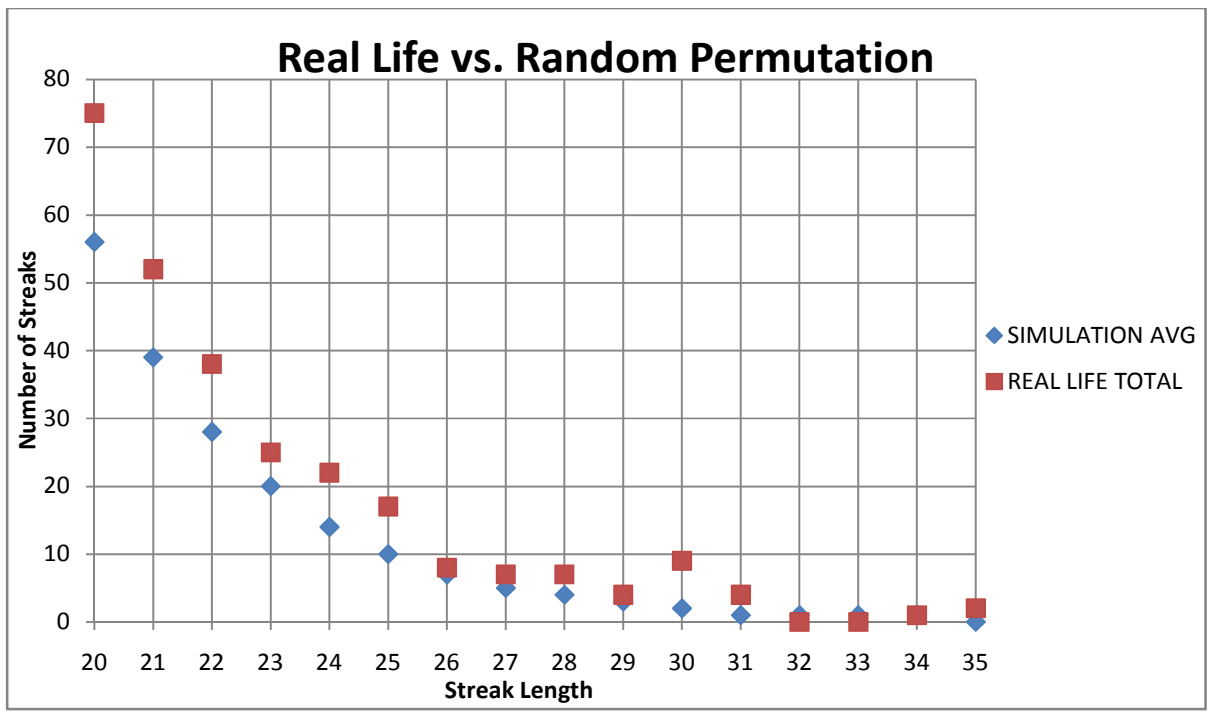
.

Notice how the number of streaks around 25 games, and especially around 30 games, spikes up, relative to the general decreasing trend of longer hitting streaks. These streaks are pretty rare, so we're dealing with small samples, but this helps show that hitters may really be paying attention to their streaks (especially their length), which lends a lot of credibility to the idea that hitters may change their behavior to keep their streaks going.

Also lending some credibility to this explanation is that the spread (the difference between how many streaks really happened and how many we *expected* to happen) seems to increase as the length of the streak increases. That is, there have been about 7 percent more hitting streaks of 10 games than we would expect, but there have been 20 percent more streaks of 15 games, and there have been 80 percent more streaks of 25 games. Perhaps, as a streak gets longer, a batter will become more focused on it, thinking about it during every at-bat, doing anything to keep it going. See the following chart for a representation of how, as streak length increases, there have been more such streaks in real life than we averaged in the random permutations.

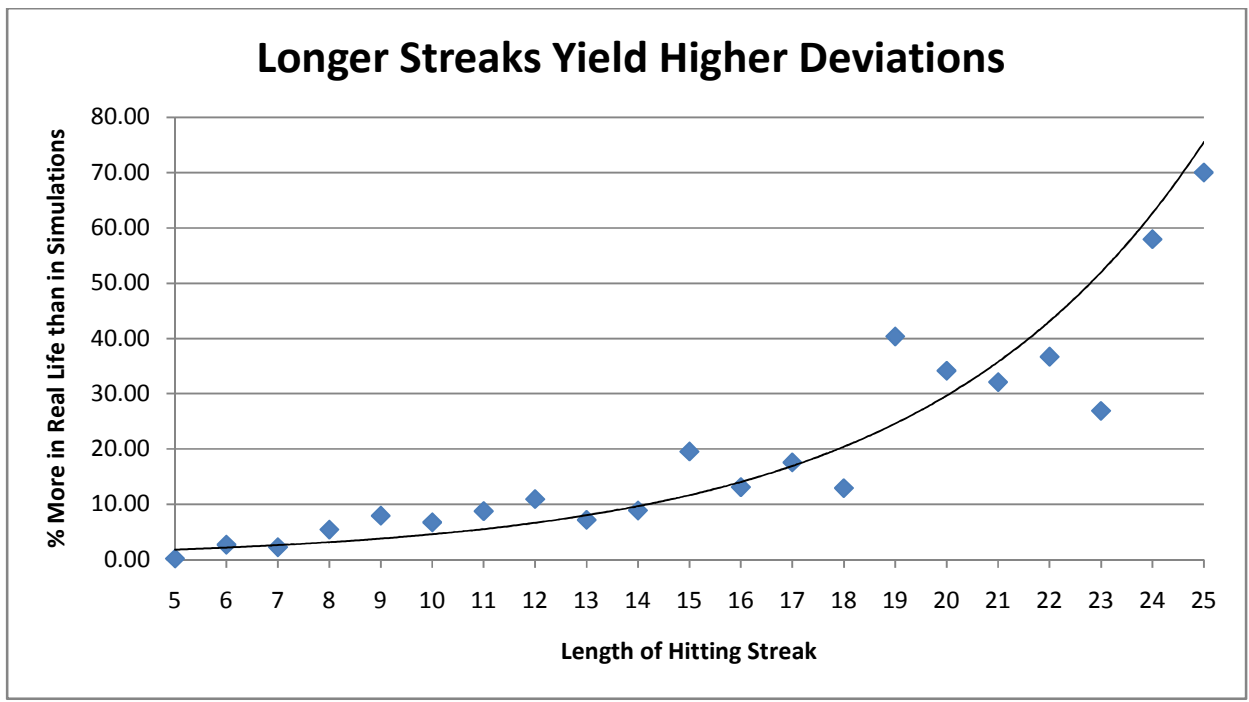
.

The evidence for this is that 85 percent of the players who had 20-plus-game hitting streaks from 1957 through 2006 had more at-bats per game *during* their hitting streak than they had for their season as a whole. Overall, it worked out to an average 6.9 percent increase in at-bats per game during their streak. That extra 6.9 percent of at-bats per game almost certainly accounts for a portion of the "extra" hitting streaks that have occurred in real life as opposed to our permutations.

This increase in at-bats per game during a streak makes sense, as batters are much less likely to be used as a pinch-hitter or be taken out of a game early when they have a hitting streak going. Additionally, when a player is hitting well, his manager is more likely to keep him in the starting lineup or even move him up in the batting order so that he gets more plate appearances. There may also be a self-fulfilling prophecy here; as a player starts hitting well, his team will tend to score more runs, which will give the batter more plate appearances. So hitting well lends itself to getting more chances to extend a hitting streak. Also, pitchers may be hesitant to walk batters (and batters hesitant to take walks) because the players want the streak to end "legitimately," with the batter being given several opportunities to extend the streak.

The extra at-bats per game also account for the slope of the previous graph, which shows an exponential trend in the number of "extra" hitting streaks that have occurred in real life as opposed to permutations. As streak length increases, those extra at-bats make streaks increasingly more likely. For instance, if we take a .350 hitter who plays 150 games and increase his at-bats per game from 4.0 to 4.28 (about a 6.9 percent increase) for an *entire season*, his odds of a 20-game hitting streak increase by 34 percent, but his odds of a 30-game streak increase by 81 percent, and his odds of a 56-game streak increase by an amazing 244 percent. Keep in mind that those increases are larger than we would see in our hitting-streak data because the 6.9 percent increase in at-bats per game applies only to the 20 or so games during the hitting streak—*not the entire 150 games that a batter plays during a season.* It is difficult to determine how many more streaks we would see if hitters' at-bats were allowed to increase by 6.9 percent for only selected stretches of their season.

**Second explanation**

Something else is going on that is significantly increasing the chances of long streaks, including possibly the idea that hitters do experience a hot-hand effect where they become more likely to have a hitting streak because they are in a period in which they continually hit better than their overall numbers suggest. This hot streak may happen at almost any point during a season, so we don't see a spike in streaks during certain parts of the year.

At first glance, the results of a hot hand would appear very similar to the hot-weather effect: If you've been hitting well lately, it's likely to continue, and if you haven't been hitting well lately, that's likely to continue as well. The difference is this: If it's the weather that's the lurking variable, then you continue hitting well because you naturally hit better during this time of year. If it's a hot-hand effect, then you continue hitting well because you're on a true hot streak. But we have seemingly shown that the weather doesn't have an effect on hitting streaks, thereby providing some credibility for the hot-hand idea.

We expect a player to have a certain amount of hot and cold streaks during any season, but the hot-hand effect says that the player will have hotter hots and cooler colds than we'd expect. So the player's overall totals still balance out, but his performance is more volatile than we would expect using the standard coin-flip model.

There may be some additional evidence for this. Over the period 1957–2006, there were about 7 percent more 3- and 4-hit games in real-life than we would expect given the coin-flip model but also about 7 percent more *hitless* games. Over a course of 50 years, those percentages really add up. What this means is that the overall numbers still balance out over the course of a season, but we're getting more "hot games" than we would expect, which is being balanced by more "cold games" than we would expect.

Additionally, there is evidence that tends to favor the hot-hand approach over the varying-at-bat approach. Dr. Mucha and I ran a second permutation of 10,000 trials that was the same as the first permutation—except we eliminated all the games where the batter did not start the game. In our first permutation, we implicitly assumed that non-starts are randomly sprinkled throughout the season. But that is likely not the case. Batters will tend to have their non-starts clustered together, usually when they return from an injury and are used as a pinch-hitter, when they have lost playing time and are used as a defensive replacement, or when they are used sparingly as the season draws to a close.

We expected that this second permutation would contain more streaks than the first permutation, as we essentially eliminated a lot of low-at-bat games, which are much more likely to end a hitting streak prematurely. The question was whether this second permutation would contain roughly the same number of streaks as occurred in *real life*.

The outcome actually comported very well with our expectations. In general, there were more streaks in this second permutation than in the first permutation—but still fewer streaks than there were in real life. For instance, in real life for 1957–2006, there were 274 streaks of 20 or more games; the first permutation (including non-starts) had an average of a mere 192 such streaks; and the second permutation (leaving out non-starts) had an average of 259 such streaks. The difference between 259 and 274 may not sound like much, but it is still very significant when viewed over 10,000 permutations, especially since we still aren't quite comparing apples to apples. There undoubtedly will be streaks that fall just short of 20 games when looking only at starts but that would go to 20 or more games when non-starts (e.g., successful pinch-hitting appearances) are included.

As the streak length increases, the difference between real life and the two permutations widens even further. For streaks of 30 or more games, there were 19 in real life, with an average of only 10 in our second permutation when we look only at starts. In this paper I deal primarily with long streaks, but I will point out that, for streaks less than 15 games, the pattern does not hold; there were fewer short streaks in real life than in the second permutation when we look only at starts.

The reason this favors the hot-hand effect is this: Our first explanation above relies on the idea that players are getting significantly more at-bats per game during their hitting streak than during the season as a whole. But the reason for a large part of that difference is that players are not frequently used as non-starters (e.g., pinch-hitters) during their streak, so it artificially inflates the number of at-bats per game that the batters get during their streaks relative to their season as a whole. Pinch-hitting appearances have little effect on real-life hitting streaks because managers are hesitant to use a batter solely as a pinch-hitter if he is hitting well. So we should be able to remove the pinch-hitting appearances from our permutations and get results that closely mirror real life. But when we do that, we still get the result that there have been significantly more hitting streaks in real life than there "should have been." This tends to add some weight to the hot-hand effect, since it just does not match up with what we would expect if the varying number of at-bats per game were the true cause.

Besides the hot-hand effect, other conditions that may be immeasurable could be playing a part. For instance, scorers may be more generous to hitters who have a long streak going, hating to see a streak broken because of a borderline call on a play that could reasonably have been ruled a base hit.

**Conclusion**

If you take away only one thing from this article, it should be this: This study seems to provide some strong evidence that players' games are not independent, identically distributed trials, as statisticians have assumed all these years, and it may even provide evidence that things like hot hands are a part of baseball streaks. It will likely take even more study to determine whether it's hot hands, or the change in behavior driven by the incentive to keep a streak going, or some other cause that really explains why batters put together more hitting streaks than they should have, given their actual game-by-game stats. Given the results, it's highly likely that the explanation is some combination of all of these factors.

The idea that hitting streaks really could be the by-product of having the hot hand is intriguing. It will tend to chafe statisticians, who rely on that key assumption of independent, identically distributed trials in order to calculate probabilities. When we remove the non-starts that could have thrown a wrench into our first permutations—*but we still get the same results*—then it really does lend some evidence for the possibility that what has happened in real life just does not match what a "random walk" would look like.

From the overwhelming evidence of the permutations, it appears that, when the same math formulas used for coin tosses are used for hitting streaks, the probabilities they yield are incorrect; those formulas incorrectly assume that the games in which a batter gets a hit are distributed randomly throughout his season. This also means that maybe all those baseball purists have had it at least partially right all this time; maybe batters really do experience periods where their hitting is above and beyond what would be statistically expected given their usual performance.

In his review of Michael Seidel's book *Streak*, Harvard biologist Stephen Jay Gould wrote:

> Everybody knows about hot hands. The only problem is that no such phenomenon exists. The Stanford psychologist Amos Tversky studied every basket made by the Philadelphia 76ers for more than a season. He found, first of all, that probabilities of making a second basket did not rise following a successful shot. Moreover, the number of "runs," or baskets in succession, was no greater than what a standard random, or coin-tossing, model would predict.

Gould's point is that hitting streaks are analogous to the runs of baskets by the 76ers in that neither should show any signs of deviating from a random coin-tossing model. I hate to disagree with a Harvard man, but my study of long hitting streaks for 1957 through 2006 seems to show that the actual number of long hitting streaks are in fact *not* the same as what a coin-tossing model would produce, even when we try to account for the fact that players get varying numbers of at-bats per game. By using the coin-flip model all of these years, we have been underestimating the likelihood that a player will put together a 20-, 30-, or even a magical 56-game hitting streak.

But this study doesn't just look at the statistic side of baseball. It also reveals the psychology of it. This study shows that sometimes batters really may have a hot hand, or at least that they adapt their approach to try to keep a long hitting streak going—and baseball players are nothing if not adapters.

---

**Coin-Flip Example**

I flipped a coin ten times and wrote down the result. I then had my computer give me a random number that is somewhere between 0 and 1, and I assigned that number to each coin flip:

| Flip Number | Result | Random Number |
|---|---|---|
| 1 | heads | 0.975 |
| 2 | tails | 0.823 |
| 3 | tails | 0.434 |
| 4 | heads | 0.191 |
| 5 | heads | 0.652 |
| 6 | tails | 0.239 |
| 7 | heads | 0.303 |
| 8 | heads | 0.009 |
| 9 | tails | 0.917 |
| 10 | heads | 0.541 |

We can consider the table above to be like John Dice's batting log. Each game with a "heads" is a game where he got a hit. Each game with a "tails" is one in which he went hitless. The longest streak of heads was two in a row.

Now, I take those results above and sort them by that random number instead:

| Flip Number | Result | Random Number |
|---|---|---|
| 8 | heads | 0.009 |
| 4 | heads | 0.191 |
| 6 | tails | 0.239 |
| 7 | heads | 0.303 |
| 3 | tails | 0.434 |
| 10 | heads | 0.541 |
| 5 | heads | 0.652 |
| 2 | tails | 0.823 |
| 9 | tails | 0.917 |
| 1 | heads | 0.975 |

It's still the same outcome as before, except that they've just been reordered completely randomly. Our longest streak of heads here is two in a row, as well. It just so happens that we end up with the same longest streak of heads in this random sorting as we did in the original tossing. But now that the results are sorted randomly, any variation in the streaks we find will be due completely to chance.

For coin tosses, we expect to find about the same number of long streaks from one trial to the next. And if hitters' results were like coin tosses, we would expect to find about the same number of long hitting streaks from one trial to the next. But my results show that the original order of baseball games (analogous to the first table of coin flips) is significantly more likely to contain long hitting streaks than the random order of baseball games (analogous to the second table of coin flips).

---

**Acknowledgments**

Peter Mucha of the Mathematics Department at the University of North Carolina deserves major applause for his great willingness to review my article and especially for writing the code that would randomly permute fifty years' worth of information a mind-boggling 10,000 times—and then doing it again for our second permutation. Had I done that same work using my original method, it would have taken me about 55 days of nonstop number-crunching. Additionally, Dr. Mucha's efforts on my project were supported in part by the National Science Foundation (award number DMS-0645369). Pete Palmer also deserves a hand for his willingness to compile fifty years of data that was essential to running my second permutation. I would be remiss if I didn't thank all of the volunteers who do work for RetroSheet, whose data made up 100 percent of the information I used in this study; Tom Ruane deserves credit for using RetroSheet data to compile several important files that contained hard-to-find information that I needed for this study. I would also like to thank Chuck Rosciam for reviewing my article, Dr. Alan Reifman of Texas Tech (who runs *The Hot Hand in Sports* blog at thehothand.blogspot.com) for reading through a preliminary copy of the article, and especially Steve Strogatz and Sam Arbesman of Cornell for offering incredible insight on this topic, for sharing their research with me, and for letting me borrow part of their *New York Times* article.

The previous article was published in the 2008 SABR Baseball Research Journal.

The next article was published in the 2009 Journal as a response and clarification.

CHARLIE PAVITT'S CRITIQUE:

I am writing this is response to Trent McCotter's piece on hitting streaks from the 2008 Baseball Research Journal. I want to begin by commending Trent on this fine piece of work. In short, a series of Monte Carlo tests revealed that the number of actual hitting streaks of lengths beginning with 5 games and ending with 35 games or more between 1957 and 2006 was, in each case, noticeably greater than what would have been expected by chance. It is always good to see evidence inconsistent with our "received wisdom." What I have to say here in no way attempts to contradict his research findings. My problem is with his attempt to explain them.

Trent proposed three "common-sense" explanations for what he found. The first was that a batter might face relatively poor pitching for a significant stretch of time, increasing the odds of a long streak. But, in his words (page 64), "the problem with this explanation is that it's too short-sided; you can't face bad pitching for too long without it noticeably increasing your numbers, plus you can't play twenty games in a row against bad pitching staffs, which is what would be required to put together a long streak." He then goes on (page 65) "The same reasoning is why playing at a hitter-friendly stadium doesn't seem to work either, since these effects don't continue for the necessary several weeks in a row." His third "common-sense explanation" is that, as hitting overall is thought to be better during the warm months, hitting streaks may be more common than expected during June through August. This is because, and this is critical (page 65), "hitting streaks are exponential…a player who hits .300 for two months will be less likely to have a hitting streak than a player who hits .200 one month and .400 the next...[because]…hitting streaks tend to highly favor batters who are hitting very well, even if it's just for a short period." This is absolutely correct. Unlike the first two proposed explanations, in this case Trent looked for relevant evidence, claiming that he looked for more streaks in June, July or August and found no more than in May. Trent, how about April and September?

Anyway, rejecting all three of these, Trent then proposed two possible psychological explanations. The first is that hitters aware of a streak intentionally change their approach to go for more singles, particularly when the streak gets long; and he has evidence that longer streaks occur less randomly than shorter ones, which would occur under this assumption (players would more likely think about keeping their streak going when it was long ongoing). The second is that hot hands really exist, and his claimed evidence is that taking games out of his random sample in which the player does not start increases the number of predicted hitting streaks, bringing it more in line with the number that actually occurred. Makes sense; a hitting streak is easier to maintain the more at bats one has in a game. He proposes that this could reflect real life because managers would start a player proportionally more often when he was hitting well. True, but we should keep in mind that the same statistical effect for starting games would occur whether there is a hit hand or not. In other words, I don't think his evidence is very telling.

Nonetheless, I want to say here that Trent may well be correct about either or both of these psychological explanations. But that doesn't matter. If we are serious about sabermetrics as a science, then without some pretty strong evidence we should NEVER use a psychological explanation for our data. The reason for this is that they can be used to explain anything, and given our present store of knowledge about player psychology they are impossible to evaluate.  Let us suppose that rather than finding more hitting streaks than chance would allow, Trent had found fewer.  He could then say that the reason for this is that batters crumble under the stress of thinking about the streak and perform worse than they would normally.  If Trent found no difference, he could then say that batters are psychologically unaffected by their circumstance.  The point, as esteemed philosopher Karl Popper pointed out in his now-classic 1934 book The Logic of Scientific Discovery, if a proposed explanation is impossible to disconfirm, then it is not scientific.  Again, Trent's proposals may be correct, but we can't judge them, so we should not be proposing them.

        The first three, however, can be disconfirmed, so we can take them seriously.  Trent claims to have disconfirmed the third, but we need to know about April and September.  But the real issue I have is with his dismissal of the first two, because he thought did not apply the logic in their case that he correctly applied for his "hot weather" proposal.  Let me begin with the first.  A batter does not have to face a bad pitching staff in consecutive games for his odds of a hitting streak to increase.  Let us suppose that a batter faces worse pitching than average during only 10 of 30 games in May and makes up for it by facing worse pitching than average during 20 of 30 games in June.  We use the same exact logic that Trent used correctly for the "hot weather" proposal; his odds of having a batting streak, which would occur during June, would be greater than another batter that faced worse pitching than average during 15 games in May and 15 games in June.  The same explanation goes for hitter-friendly and -unfriendly ballparks, and is strengthened in this case because of well-supported known differences in ballpark effects.  If a player's home field was hitter-friendly and, during a stretch of time, many of his road games were in hitters' parks, he could easily have 20 or more games in this context in a given month.

        I have no idea whether either of these two explanations for Trent's findings is correct.  But the difference between these and his psychological proposals is that we could test these two and not those he favors.  I would obviously like to see that happen.  And I would very much like it if we dispense with psychological explanations for our findings until we are in a position to evaluate them.

McCotter response to Pavitt's review of 'Hitting Streaks Don't Obey Your Rules.'

I'm glad that my article, "Hitting Streaks Don't Obey Your Rules," has fostered some debate on the topic of hitting streaks.  Charlie Pavitt has written an excellent review, and I have also received numerous e-mails from others who read the article and had insights and critiques.

Below, I have tried to outline some rebuttals, clarifications, and corrections concerning the article.

First, I want to state that the original intent of my article was to try to disprove the standard independence assumption that underlies the coin-toss model that is used to calculate probabilities of streaks.  For a long time now, players' final season statistics (like 150 games, 600 at-bats, 200 hits) were used to extrapolate what kinds of hitting streaks were likely to have happened.  It works fine, so long as the games are essentially randomly distributed.  By randomly permuting the games 10,000 times for each player over 1957-2006—and getting so many fewer streaks than we have seen in real life—I think there is very strong evidence that the independence assumption underlying the coin-toss model does not work in the context of hitting streaks.

Once I had shown the poor results of the independence test, I tried to come up with explanations for why we have seen so many more hitting streaks than occurred in the random permutations.

I'll now address several of Charlie Pavitt's arguments on that topic.   He says that I've too-quickly dismissed the effects of facing bad opponents.  It's very difficult to test the effect of facing bad pitching because there is no easy way to define bad pitching.  One short-cut I have used is to look at how many long hitting streaks there have been against particular teams (i.e., a batter getting a hit in 30 straight games vs. the Blue Jays, over the course of his career).  Over 1957-2006, there have been 19 hitting streaks of 30 or more games vs. the league as a whole, but only 5 such streaks vs. a particular opponent.  We expect fewer streaks, simply because you can't count the last 10 games vs. Toronto and the first 20 games vs. Texas as a 30-game hitting streak vs. one particular opponent.  But if facing bad teams were so conducive to hitting streaks, it seems like we would have seen more hitting streaks against bad teams—those bad teams would continually be boosting their opponents' averages.

Pavitt also says that I too-quickly dismissed the effect of playing at certain hitter-friendly ballparks.  I went through all 19 of the 30-game hitting streaks over 1957-2006.  Over those streaks, 50.2% of the games comprising the hitting streaks were played at the batters' home stadiums, and 49.8% of the games were played at road stadiums.  Batters get more at-bats when they're on the road (since their team always gets to bat in the ninth inning), but batting averages are higher at home stadiums.  In the end, it clearly has balanced out.  Thus, players who have had 30-game hitting streaks don't seem to have received any advantage either by playing more games at home (where they have a higher batting average), or by playing more games on the road (where they get more at-bats).  If playing at a hitter-friendly stadium greatly helped long hitting streaks, it seems like we'd see a higher percentage than 50.2% of the games making up the streak to have been played at the hitter-friendly home park.  In other words, long hitting streaks over 1957-2006 don't seem to be centered around stretches where the player was playing more games at home or on the road than they do at any given stretch of the season.

Pavitt also mentions that I didn't include any data about hitting streaks beginning in April or September. I exclude April because the season's beginning date frequently changed over 1957-2006, and it often wasn't until mid-April. So there just weren't as many games being played in April as there were in May, June, July, or August. I exclude September because streaks that begin in September have a much lower chance of actually making it to 20 or 30 games, simply because the player will run out of games to play. So comparing April or September to the months of May, June, July, and August wouldn't give us any insight.

Pavitt's final critique is that several of my 'psychological' explanations for extra hitting streaks aren't testable. I agree that testing these things is very difficult, simply because that's the nature of testing humans, who can adjust on-the-fly. However, just because something may be difficult—or even impossible—to verify doesn't mean that we should exclude it as a possible factor. For instance, the placebo effect with drugs is a psychological explanation that seems very difficult to refute; we accept it as valid mostly because we've eliminated other explanations. In the baseball world, there is a common thought that batters tend to take fewer walks as their hitting streaks increase. This is tough to test, but not impossible; maybe it's just a result of multiple effects that naturally vary at-bats throughout the season. I just don't see the problem with including psychological factors in a study that deals with human behavior.

I also wanted to include some quick answers to questions that repeatedly came up:

1) I only looked at single-season streaks for the entire project, so multi-season streaks are not included.

2) I excluded all 0-for-0 batting lines, EXCEPT where the player had a sacrifice fly. Per the MLB Rules, an 0-for-0 with a sacrifice fly will end the hitting streak, even though the batter had zero at-bats.

And here is one correction to the original article: the y-axis of the chart on page 66 should read from 0% to 80%, instead of 0% to 8%. Thus, for example, we saw an extra 40%—not 4%—in 18-game hitting streaks.

streak attrition.txt

If all you told me was that a player had a 20+ game hitting streak, I'd be able to tell you that the most-likely place it'd end would be when the player was trying to go from game #30 to game #31.

Using these totals I've compiled (which may be a little off, since there are still hitting streaks out there waiting to be found), moving from the 30th game to the 31st game has the lowest success rate out of all sequences of at least 20 games in a row (at least til you hit 45 games, where only 50% (DiMaggio but not Keeler) went from 45 to 46 games).

Check out these success rates for continuing a streak from length 'n' to 'n+1':

of all streaks that hit 20g, 75.3% of them continued to at least 21g
of all streaks that hit 21g, 75% of them continued to at least 22g
"" 22 games, 77.1% continued to at least 23g
23g->24+g: 74.5%
24g->25+g: 79.8%
25g->26+g: 77.3%
26g->27+g: 75.8%
27g->28+g: 77.6%
28g->29+g: 77.8%
29g->30+g: 74.3%
30g->31+g: 64.2% (34-out-of-53)
31g->32+g: 70.6% (24-out-of-34)
32g->33+g: 91.7%
33g->34+g: 81.8%
34g->35+g: 83.3%
35g->36+g: 66.7% (10-out-of-15)
36g->37+g: 90%
37g->38+g: 88.9%
38g->39+g: 87.5%
39g->40+g: 85.7%
40g->41+g: 83.3%

...the longer streaks suffer from smaller sample sizes (the last sequence where we have at least 20 'samples' is moving from 33 to 34 games, where 18-out-of-22 players made it).

But I assume you can pick out the real outlier: of all streaks that hit 30 games, only 64.2% of them move on to 31 games, the lowest success rate for moving from any length hitting streak to the next game (except, as mentioned above, for the really long streaks where there have been so few).

So, given that Zimmerman's streak was more than 20 games (but less than 46 games), it was more-likely to fail at game #31 than at any other game.

Trent McCotter

| 1957-2006 | ACTUALLY HAPPENED | IN THE 10,000 RANDOM SORTINGS | |
|---|---|---|---|
| LENGTH | # OF HITTING STREAKS | AVG | STAN. DEV. |
| 5 | 22632 | 22584.63 | 141 |
| 6 | 14470 | 14086.60 | 112 |
| 7 | 9151 | 8947.64 | 89 |
| 8 | 6081 | 5766.29 | 72 |
| 9 | 4059 | 3759.81 | 59 |
| 10 | 2645 | 2477.50 | 48 |
| 11 | 1792 | 1647.42 | 39 |
| 12 | 1226 | 1104.86 | 32 |
| 13 | 801 | 747.12 | 27 |
| 14 | 552 | 506.85 | 22 |
| 15 | 415 | 347.13 | 19 |
| 16 | 270 | 238.69 | 16 |
| 17 | 194 | 164.97 | 13 |
| 18 | 129 | 114.22 | 11 |
| 19 | 112 | 79.80 | 8.9 |
| 20 | 75 | 55.90 | 7.5 |
| 21 | 52 | 39.36 | 6.2 |
| 22 | 38 | 27.80 | 5.3 |
| 23 | 25 | 19.70 | 4.4 |
| 24 | 22 | 13.93 | 3.7 |
| 25 | 17 | 10.00 | 3.2 |
| 26 | 8 | 7.20 | 2.7 |
| 27 | 7 | 5.13 | 2.3 |
| 28 | 7 | 3.71 | 1.9 |
| 29 | 4 | 2.63 | 1.6 |
| 30 | 9 | 1.90 | 1.4 |
| 31 | 4 | 1.39 | 1.2 |
| 32 | 0 | 1.01 | 0.99 |
| 33 | 0 | 0.74 | 0.86 |
| 34 | 1 | 0.55 | 0.74 |
| 35+ | 5 | 1.48 | 1.21 |

| 1957-2006 | ACTUALLY HAPPENED | IN THE 10,000 RANDOM SORTINGS | | |
|---|---|---|---|---|
| LENGTH | # OF HITTING STREAKS | AVG | STAN. DEV. | PROB. |
| 5+ | 64803 | 62765.96 | 150.69 | 6.4 * 10^-42 |
| 10+ | 8410 | 7620.99 | 69.45 | 3.4 * 10^-30 |
| 15+ | 1394 | 1137.24 | 30.71 | 3.2 * 10^-17 |
| 20+ | 274 | 192.43 | 13.32 | 4.6 * 10^-10 |
| 25+ | 62 | 35.74 | 5.84 | 3.5 * 10^-6 |
| 30+ | 19 | 7.07 | 2.60 | 2.2 * 10^-6 |
| 35+ | 5 | 1.48 | 1.21 | 0.0018 |

| 1957-2006 | ACTUALLY HAPPENED | 10,000 SORTINGS--*STARTS ONLY* | |
| --- | --- | --- | --- |
| LENGTH | # OF HITTING STREAKS | AVG | STAN. DEV. |
| 5 | 22632 | 22557 | 141 |
| 6 | 14470 | 14743 | 115 |
| 7 | 9151 | 9721 | 93.1 |
| 8 | 6081 | 6466 | 76.6 |
| 9 | 4059 | 4335 | 62.8 |
| 10 | 2645 | 2920 | 52.2 |
| 11 | 1792 | 1979 | 43.3 |
| 12 | 1226 | 1349 | 35.8 |
| 13 | 801 | 924 | 29.8 |
| 14 | 552 | 636 | 25.1 |
| 15 | 415 | 440 | 20.5 |
| 16 | 270 | 306 | 17.2 |
| 17 | 194 | 213 | 14.5 |
| 18 | 129 | 149 | 12.2 |
| 19 | 112 | 105 | 10.0 |
| 20 | 75 | 74.2 | 8.54 |
| 21 | 52 | 52.5 | 7.20 |
| 22 | 38 | 37.2 | 5.97 |
| 23 | 25 | 26.6 | 5.18 |
| 24 | 22 | 19.0 | 4.35 |
| 25 | 17 | 13.7 | 3.68 |
| 26 | 8 | 9.82 | 3.14 |
| 27 | 7 | 7.04 | 2.64 |
| 28 | 7 | 5.10 | 2.25 |
| 29 | 4 | 3.70 | 1.92 |
| 30 | 9 | 2.68 | 1.62 |
| 31 | 4 | 1.95 | 1.40 |
| 32 | 0 | 1.44 | 1.21 |
| 33 | 0 | 1.05 | 1.03 |
| 34 | 1 | 0.77 | 0.87 |
| 35+ | 5 | 2.19 | 1.47 |

| 1957-2006 | ACTUALLY HAPPENED | 10,000 SORTINGS--*STARTS ONLY* | | |
| --- | --- | --- | --- | --- |
| LENGTH | # OF HITTING STREAKS | AVG | STAN. DEV. | PROB. |
| 5+ | 64803 | 67102 | 151.00 | 1.0000 |
| 10+ | 8410 | 9280 | 75.50 | 1.0000 |
| 15+ | 1394 | 1472 | 34.40 | 0.9880 |
| 20+ | 274 | 259.00 | 15.20 | 0.1620 |
| 25+ | 62 | 49.40 | 6.88 | 0.0336 |
| 30+ | 19 | 10.10 | 3.12 | 0.0022 |
| 35+ | 5 | 2.19 | 1.47 | 0.0281 |